\documentclass[submission,copyright,creativecommons]{eptcs}
\providecommand{\event}{ICE 2014} 

\usepackage{listings}
\usepackage{etex}

\newcommand{\checkDtmc}{\textsf{Check}}
\newcommand{\checkPath}{\textsf{CheckPath}}
\newcommand{\checkBoundedUntil}{\textsf{CheckBoundedUntil}}
\newcommand{\checkUnBoundedUntil}{\textsf{CheckUnboundedUntil}}

\newcommand{\relop}{\bowtie}

\usepackage{tikz}
\usetikzlibrary{calc,arrows,shapes,chains,matrix,positioning,scopes,decorations.pathmorphing,shadows,fit,mindmap}
\definecolor{rosso}{rgb}{.99,.5,.5}
\definecolor{dgreen}{rgb}{0,.7,0}

\usepackage{pgf}
\usepackage{stmaryrd,amsmath,amsfonts,amstext,amssymb,mathrsfs}
\usepackage{eufrak}
\usepackage{arrows}

\usetikzlibrary{matrix}
\usetikzlibrary{arrows}
\usetikzlibrary{automata}

\usepackage[algoruled,linesnumbered,lined,english]{algorithm2e}

\usepackage{color}
\usepackage{colortbl}
\usepackage{mathrsfs}
\definecolor{apricot}{RGB}{249,243,221}
\def\RED#1{\textcolor{red}{#1}}

\newcommand{\defdtmc}{
(S,\overline{s},\mathbf{P})
}

\newcommand{\nxt}{{\mathsf{next}}}

\newcommand{\lab}{{\mathsf{lab\_eval}}}

\usepackage{breakurl}
\usepackage{amstext,amsmath,amssymb}
\usepackage{latexsym}
\usepackage{textcomp}
\usepackage{subfigure}
\usepackage{graphicx}

\usepackage[arrow,matrix,frame,curve,cmtip]{xy}

\newtheorem{theorem}{Theorem}[section]
\newtheorem{lemma}{Lemma}[section]

\newcommand{\qed}{}

\newcommand{\AND}{\mbox{$\;\;\wedge\;\;$}}

\newcommand{\All}{\mbox{$\forall\,$}}
\newcommand{\Ex}{\mbox{$\exists\,$}}

\def\Prob#1#2#3{{\calP}_{#1 #2}(#3)}

\def\rU#1{\, {\calU}^{#1} \,}

\def\calM{{\cal M}}

\def\calP{{\cal P}}

\def\calR{{\cal R}}
\def\calS{{\cal S}}

\def\calU{{\cal U}}

\def\calX{{\cal X}}

\renewcommand{\iff}{\mbox{ iff }}

\newcommand{\Paths}{\mbox{\sl Paths}}

\renewcommand{\Pr}{{\mathbb{P}}}

\def\calS{{\cal S}}

\def\lnxt{\calX \,}			

\def\slf{\ell}			
\def\sls{\mathscr{P}}			
\def\sdtm{\mathbf{P}}	

\newtheorem{nfa}{\RED{Note for Authors}}[section]

\newif\ifTR
\TRtrue

\begin{document}

\hyphenation{System-of-In-de-pend-ent}


\title{On-the-fly Probabilistic Model Checking}

\pagestyle{plain}


\author{Diego Latella
\institute{ISTI - CNR}
\and
Michele Loreti
\institute{Universit\`a di Firenze}
\and
Mieke Massink
\institute{ISTI - CNR}
}

\def\titlerunning{On-the-fly Probabilistic Model Checking}
\def\authorrunning{D. Latella, M. Loreti \& M. Massink}
\maketitle

\noindent

\ifTR
\begin{abstract}
Model checking approaches can be divided into two broad categories: global approaches that determine the set of all states in a model $\calM$ that satisfy a temporal logic formula $\Phi$, and local approaches in which, given a state $s$ in $\calM$, the procedure determines whether s satisfies $\Phi$. When $s$ is a term of a process language, the model-checking procedure can be executed ``on-the-fly'', driven by the syntactical structure of $s$.
For certain classes of systems, e.g. those composed of many parallel components, the local approach is preferable because, depending on the specific property, it may be sufficient to generate and inspect only a relatively small part of the state space.
We propose an efficient, on-the-fly, PCTL model checking procedure that is parametric with respect to the semantic interpretation of the language. The procedure comprises both bounded and unbounded until modalities. The correctness of the procedure is shown and its efficiency is compared with a global PCTL model checker on representative applications.
\end{abstract}
\fi

\section{Introduction and Related Work}
\label{Introduction}

\newcommand\FlyFast[1]{{\sf FlyFast#1}}


Model checking approaches are often divided into two broad categories: {\em global} approaches that determine the set of all states in a model $\calM$ that satisfy a temporal logic formula $\Phi$, and {\em local} approaches in which, given a state $s$ in $\calM$, the procedure determines whether s satisfies $\Phi$~\cite{Courcoubetis1992,BCG95}. When $s$ is a term of a process language, the model checking procedure can be executed ``on-the-fly'', driven by the syntactical structure of $s$. On-the-fly algorithms are following a {\em top-down} approach that does not require global knowledge of the complete state space. For each state that is encountered, starting from a given state, the outgoing transitions are followed to adjacent states, constructing step by step local knowledge of the state space until it is possible to decide whether the given state satisfies the formula (or memory bounds are reached).
Global algorithms instead, construct the set of states that satisfy a formula recursively in a {\em bottom-up} fashion following the syntactic structure of the formula~\cite{Clarke1986} and require the full state space of the model to be generated before they can be applied.

In this paper, we present a local, on-the-fly, probabilistic model checking algorithm for full \emph{Probabilistic Computation Tree Logic} (PCTL)~\cite{Hansson1994}, a probabilistic extension of the temporal logic \emph{Computation Tree Logic} (CTL)~\cite{Clarke1986} that includes both the bounded and unbounded until operator. The algorithm is {\em parametric} with respect to the semantic interpretation of the front-end language. Each instantiation of the algorithm consists of the appropriate definition of two functions: $\nxt$ and $\lab$. Function $\nxt$, given a process term (or state)\footnote{We will consider process terms as states throughout this paper.}, returns a list of pairs. Each pair consists of a process term, that can be reached in one step from the given process term, and its related probability.
Function $\lab$, given a term, gives a boolean function associating {\sf true} to each atomic proposition with which the term is labelled.
This parametric approach has the advantage that the model checker can be easily instantiated on specification languages with different semantics. For example, in~\cite{La+13a} we present two different interpretations for bounded PCTL; one being the standard, exact probabilistic semantics of a simple, time-synchronous population description language, and the other being the mean-field approximation in discrete time of such a semantics~\cite{BMM07,La+13a,LLM13b}. The mean-field approximation has proved a successful technique to analyse properties of individual components in the context of large population models in the discrete time setting\footnote{The mean-field technique approximates the mean global behaviour of the population by a deterministic limit that provides at each time step the expected number of objects that are in the various local states. The iterative calculation of the mean-field in combination with the process modelling the single object forms a DTMC that lends itself very well to on-the-fly analysis and the computational complexity is insensitive to the size of the population.}. The main contributions of the current paper are twofold: 1) we provide a detailed description of the on-the-fly algorithm (not presented in~\cite{La+13a,LLM13b}) together with the proofs of correctness. In particular, the algorithm for the unbounded until operator uses a new technique exploiting an interesting property of transient Discrete Time Markov Chains (DTMCs), i.e. one in which all recurrent states are absorbing; 2) we use an instantiation of the prototype on-the-fly PCTL model checker \FlyFast\ on an automata based language and semantics such as that used in the PRISM model checker which in turn is used to make a first comparison for what concerns the efficiency of the on-the-fly algorithm when used for full PCTL model checking to make sure that its efficiency is at least comparable with PRISM in the worst case in which the whole state space must be explored. 



\noindent
\textbf{Related work.}
%
%
In the context of qualitative model checking of temporal logics such as CTL ~\cite{Clarke1986}, LTL~\cite{Pnueli1977,RoV10} and CTL*\cite{BCG95}, local model checking algorithms have been proposed to mitigate the state space explosion problem using an on-the-fly approach~\cite{Courcoubetis1992,BCG95,Hol04,GnM11}. 
The have also the same worst-case complexity as the best existing global procedures for the above mentioned logics. However, they have better performance  when only a subset of the system states need to be analysed to determine whether a system satisfies a formula. Such cases occur frequently in practice. Furthermore, local model checking may provide results for infinite state spaces.

In the context of probabilistic and stochastic model checking global algorithms have been more popular than local ones and can be found in many sophisticated tools such as PRISM, MRMC and many others~\cite{Ba+03,KNP04}. 
A clear advantage of these global algorithms is that results are obtained for {\em all} states of the model, if the state space is not too large, and that, depending on the particular formula to verify, usually the underlying model can be reduced to fewer states before the algorithm is applied and 
can be reduced to combinations of existing well-known and optimised algorithms for Markov chains such as transient analysis~\cite{Ba+03}. In the context of Markov Decision Processes (MDP) partial order reduction techniques have been explored to obtain state space reduction~\cite{FBBF12}. This technique is based on a static partial order reduction approach that, starting from 
the complete state space representation, produces an equivalent and compact representation of the state space
that can be used as input of the model checking algorithm~\cite{BDG06}.

To the best of our knowledge the only algorithm for on-the-fly model checking for probabilistic processes is the
one proposed in~\cite{DIMTV04} that only considers the fragment of the
PCTL without unbounded until. On the contrary, the local model-checking algorithm considered in this paper 
considers all PCTL. This is an important point. Indeed, the use of full PCTL forbids the application of specific 
techniques, like for instance statistical model checking, e.g. \cite{Younes2004}, that can be used only when 
computations with a bounded temporal horizontal. 
%
%
%
%
%
%

\section{Probabilistic Computation Tree Logic}
\label{sec:preliminaries}
\label{Logic}

In this section we  briefly recall the definition of the \emph{Probabilistic Computation Tree Logic} (PCTL)~\cite{Hansson1994}, a probabilistic extension of the temporal logic CTL~\cite{Clarke1986}, for the expression of properties of Discrete Time Markov Chains (DTMCs) and Markov Decision Processes (MDPs).
The syntax of PCTL is the following:
\[
\Phi  ::=  a \mid \neg \, \Phi \mid  \Phi \, \vee \, \Phi \mid \Prob{\bowtie}{p}{\varphi} \qquad
\mbox{ where }
\varphi ::=  \lnxt \Phi \mid  \Phi \rU{\le k} \Phi \mid  \Phi \rU{} \Phi 
\]
%
%
where $a \in \sls$ is an atomic proposition, $\bowtie \; \in \{ \leq,<,>,\geq\}$, $p \in [0,1]$ and $k \in \mathbb{N}$. 
PCTL formulas are interpreted over {\em state labelled}  DTMCs and consist of all the state formulas $\Phi$. The path formulas $\varphi$ only appear as parameter of the operator $\Prob{\bowtie}{p}{\varphi}$.  Informally,
a state $s$ in a DTMC satisfies $\mathcal{P}_{\bowtie p}[\varphi]$ if the total probability measure of the set of paths that satisfy path formula $\varphi$ is $\bowtie p$. 
A state labelled DTMC is a pair $\langle \calM, \slf  \rangle$ where $\calM$ is
a DTMC with state set $\calS$ and $\slf : \calS \rightarrow 2^\sls$ associates each state
with a set of atomic propositions; for each state $s \in \calS$, $\ell(s)$ is the
set of atomic propositions true in $s$. In the following, we assume $\sdtm$ 
be the one step probability matrix for $\calM$; we abbreviate
$\langle \calM, \slf  \rangle$ with $\calM$, when no confusion can arise.
A path $\sigma$ over  $\calM$ is a non-empty sequence of states
$s_0, s_1, \cdots$ where  $\sdtm_{s_i,s_{i+1}} >0$
for all $i\ge 0$. 
We let $\Paths_{\calM}(s)$ denote the set of all infinite paths over $\calM$ starting from state $s$. 
By $\sigma[i]$ we denote the $i$-th  element $s_i$ of path 
$\sigma$, for $i \geq 0$.
The satisfaction relation on $\calM$ and the logic are formally defined in Table~\ref{D:DPCTLSAT}.
For every path formula $\varphi$, the set $\left\{\sigma \in \Paths_{\calM}(s) | \sigma \models \varphi \right\}$ is a \emph{measurable set}~\cite{KNP04}.
\begin{table}[tbp]
\begin{center}
\fbox{
\begin{minipage}{4.5in}
\footnotesize
$
\begin{array}{lcl}
s \models_{\calM} a & \iff & a \in \ell(s) \\[1ex]
s \models_{\calM} \neg \Phi & \iff & \mbox{not } s \models_{\calM} \Phi \\[1ex]
s \models_{\calM} \Phi_1 \, \vee \, \Phi_2 & \iff & s \models_{\calM} \Phi_1 
\mbox{ or } s \models_{\calM} \Phi_2\\[1ex]
s \models_{\calM} \Prob{\bowtie}{p}{\varphi} & \iff &
   \Pr \{ \sigma \in \Paths_{\calM}(s) \mid \sigma \models_{\calM} \varphi \} \bowtie p\\[0.20cm]
\sigma \models_{\calM} \, \lnxt \, \Phi & \mbox{iff  } & \sigma[1] \models_{\calM} \Phi \\
\sigma  \models_{\calM} \Phi_1 \rU{\le k}{} \Phi_2 & 
\mbox{iff  } & 
\Ex 0 \le h \le k \;s.t.\; \sigma[h] \models_{\calM} \Phi_2 \AND 
\All 0\le i <  h \;.\;  \sigma[i] \models_{\calM} \Phi_1\\
\sigma  \models_{\calM} \Phi_1 \rU{}{} \Phi_2 & 
\mbox{iff  } & 
\Ex 0 \leq k \;s.t.\; \sigma[k] \models_{\calM} \Phi_2 \AND 
\All 0\leq i <  k \;.\;  \sigma[i] \models_{\calM} \Phi_1\\
\end{array}
$
\normalsize
\end{minipage}
}
\end{center}
\caption{Satisfaction relation for PCTL.\label{D:DPCTLSAT}}
\end{table}

\section{On-the-fly Probabilistic Model Checking}
\label{PCTLaMC}


We introduce a local on-the-fly model checking algorithm for PCTL on labeled DTMC
$\langle \calM, \slf  \rangle$. The basic 
idea of an on-the-fly algorithm is simple: while the state space is generated in a stepwise fashion from a term $s$ of the language, the algorithm keeps track of all 
the paths that are being generated. For each of them it updates the information about the satisfaction of the formula 
that is checked. In this way, only that part of the state space is generated that may provide information on the satisfaction 
of the formula and irrelevant parts are not taken into consideration, mitigating the problem of  \emph{state space} explosion. 
%
However, the proposed model checking algorithm is not only based on graph generation. 
Indeed, while the relevant part of the state space is generated, the satisfaction probabilities of path formulas are also computed (on-the-fly). 


The proposed algorithm abstracts from any specific language
and from different semantic interpretations of a language. We only assume an abstract interpreter function 
that, given a generic process term, returns a probability distribution over the set  of terms. 
Below, we let \textsf{proc} be the (generic) type of \emph{probabilistic process terms}\footnote{}
while we let \textsf{formula} and \textsf{path\_formula} be the types of \emph{state-} and \emph{path-}
PCTL formulas. Finally, \textsf{lab} denotes the type of \emph{atomic propositions}.

The abstract interpreter can be modelled by means of two functions: 
$\nxt$ and $\lab$.
Function $\nxt$ associates a list of pairs $(\mathsf{proc},\mathsf{float})$ to each element of type \textsf{proc}. The list of pairs gives the terms, i.e. states, that can be reached in one step from the given state and their one-step transition probability. We require that for each $s$ of type $\mathsf{proc}$ it  holds that
$0 < p' \leq 1$,  for all $(s',p') \in \nxt(s)$ 
and
$ { \sum_{(s',p')\in \nxt(s)} p'= 1}$.
Function $\lab$ returns for each element of type \textsf{proc} a function associating a \textsf{bool} to each atomic proposition $a$ in $\textsf{lab}$.
Each instantiation of the algorithm consists in the appropriate definition
of $\nxt$ and $\lab$, depending on the language at hand and its semantics.

The local model checking algorithm is defined as a function, \checkDtmc, 
shown in Table~\ref{alg:check_dtmc}.
On atomic state-formulas, the function returns the value
of $\lab$; when given a non-atomic state-formula, \checkDtmc{}  
calls itself recursively on sub-formulas, in case they are state-formulas, whereas it calls function 
\checkPath, in case the sub-formula is a path-formula.
In both cases the result is a Boolean value that indicates whether the state satisfies the formula\footnote{For obvious reasons of presentation here we show a simplified, not fully optimised, pseudo-code version of the algorithm.}.
\begin{table}[tbp]
\begin{lstlisting}
boolean $\checkDtmc$( $s: \mathsf{proc},$  $\Phi: \mathsf{formula}$) {
  switch ($\Phi$) {
    case $a$: return $\lab(s,a)$;
    case $\neg \Phi_1$: return $\neg\checkDtmc$( $s$ , $\Phi_1$ );
    case $\Phi_1 \vee \Phi_2$: return $\checkDtmc$( $s$ , $\Phi_1$ ) $\vee$ $\checkDtmc$( $s$ , $\Phi_2$ );
    case $\Prob{\relop}{p}{\varphi}$: return $\checkPath(s,\varphi)\relop p$; 
  }
}
\end{lstlisting}
\caption{\label{alg:check_dtmc} Function $\checkDtmc$}
\end{table}

Function \checkPath, shown in Table~\ref{alg:checkpath_dtmc}, takes two input parameters: a state $s \in \textsf{proc}$ and a PCTL path-formula $\varphi \in \mbox{\textsf{path\_for\-mu\-la}}$. As a result, it produces the probability measure of the set of paths, starting in state $s$, which satisfy path-formula $\varphi$. Following the definition of the formal semantics of PCTL, three 
different cases can be distinguished. If $\varphi = \lnxt \Phi$ then the result is  the sum of the probabilities of the transitions from $s$ to those next states $s'$ that satisfy $\Phi$. To verify the latter, function \checkDtmc{} is recursively invoked on such states. If $\varphi$ is 
$\Phi_1 \rU{\le k}{} \Phi_2$ or $\Phi_1 \rU{}{} \Phi_2$ functions $\checkBoundedUntil$ or $\checkUnBoundedUntil$ are invoked, respectively. These functions are presented in the next two subsections.

\begin{table}[tbp]
\begin{lstlisting}
float $\checkPath$( $s: \mathsf{proc}$,  $\varphi: \mathsf{path\_formula}$ ) {
 switch $\varphi$ {
  case $\lnxt \Phi$: {
    $p$ = 0.0;
    $lst$ = $\nxt(s)$;
    for $(s',p')\in lst$ {
     if ($\checkDtmc( s' , \Phi )$) { $p$ = $p + p'$;}
    }
    return $p$;
  }
  case $\Phi_1 \rU{\le k}{} \Phi_2$: return $\checkBoundedUntil$( $s$ , $\Phi_1$ , $k$ ,  $\Phi_2$ );
  case $ \Phi_1 \rU{}{} \Phi_2$: return $\checkUnBoundedUntil$( $s$ , $\Phi_1$ ,  $\Phi_2$ );
 }
}
\end{lstlisting}
\caption{\label{alg:checkpath_dtmc} Function $\checkPath$}
\end{table}

Let $s$ be a term of a probabilistic process language and $\calM$ 
the complete discrete time stochastic process associated with
$s$ by the formal semantics of the language. 
\begin{theorem}
\label{LMC:CORR}
$s \models_{\calM}  \Phi$ if and only if $\checkDtmc(s,\Phi)=\mathsf{true}$.
\end{theorem}
%

\noindent
\textbf{Proof.}
The theorem is proven by induction on the structure of $\Phi$. The more involved parts concern the proof of the theorem for the path formulas concerning bounded and unbounded until. These are provided as Lemma~\ref{lemma:checkboundeduntil}, Lemma~\ref{lemma:checkunboundeduntil} and Lemma~\ref{lemma:checkunboundeduntiltermination} together with an outline of their proofs in the following sections. 

{\hfill$\Box$}

\subsection{Computing Bounded Until Probability}
\newcommand{\createBUStructure}{\mathsf{createBUStructure}}
\newcommand{\createUStructure}{\mathsf{createUStructure}}
\newcommand{\YESLabel}{\mathsf{YES}}
\newcommand{\NOLabel}{\mathsf{NO}}
\newcommand{\UNKNOWNLabel}{\mathsf{UNKNOWN}}
\newcommand{\expandBoundUntil}{\mathsf{expandBU}}
\begin{table}[t!]
\begin{lstlisting}
float $\checkBoundedUntil$($s: \mathsf{proc}$ , $\Phi_1:\mathsf{formula}$ , $k:\mathsf{int}$ ,  $\Phi_2: \mathsf{formula}$) {
 r = $\createBUStructure$( s , $\Phi_1$ , $k$ , $\Phi_2$ );
 $M$ = [$s \mapsto r$];
 if ($r.label == \YESLabel$) { return 1.0; }
 if ($r.label == \NOLabel$) { return 0.0; }
 $S_{yes}$ = $\emptyset$;
 $toExpand = \{ r \}$;
 $c=k$;
 while $(c>0)\wedge (toExpand \not= \emptyset)$ {
  $T=toExpand$;
  $toExpand = \emptyset$;
  for ($r\in T$) {
   $lst$ = $\nxt(r.term)$;
   for $(s',p')\in lst$ {
    $r'$=$M[s']$;
    if ($r'==\bot$) {
      $r'$ = $\createBUStructure$( $s'$ , $\Phi_1$ , $k$ , $\Phi_2$ );
      $M$ = $M[s' \mapsto r']$;
      if ($r'.label$ == $\YESLabel$) { 
       $S_{yes}$= $S_{yes}\cup \{ r' \};$ 
      } else if ($r'.label$ != $\NOLabel$) {
       $toExpand$ = $toExpand\cup \{ r' \}$; 
      } 
    }
    $r'.prec$ = $(r,p)::r'.prec$;
   }
  }
  $c$ = $c-1$;
 }
 if ($S_{yes}==\emptyset$) { return 0.0; }
 $A$ = $S_{yes}$;
 for ($i=1$;$i<=k$;$i++$) {
  for ($r\in A$) {
   for ($(r',p')\in r.prec$) {
    $r'.p[i]$ = $r'.p[i]+p'*r.p[i-1]$;   
   }
  }
  $A$ = $\{ r | \exists r'\in A: r\preceq r' \}$;
 }
 return $r.p[k]$;
}
\end{lstlisting}
\caption{\label{alg:check_boundeduntil} Function $\checkBoundedUntil$}
\end{table}
%
%
%
%
Function $\checkBoundedUntil$, defined in Table~\ref{alg:check_boundeduntil}, 
computes the probability of the set of paths starting from a given state $s$ that satisfy formula $\Phi_1 \rU{\le k}{} \Phi_2$. 
This function takes as parameters a state $s$, state formulas $\Phi_1$ and $\Phi_2$, and the bound $k$.
Notice that differently from the algorithm proposed in~\cite{DIMTV04}, where a recursive algorithm is proposed, 
to compute the probability of path satisfying $\Phi_1 \rU{\le k}{} \Phi_2$ an iterative solution is proposed. 
Moreover, the proposed procedure is not an adaptation of standard PCTL where all the state space is considered to
compute the requested probability value.

To compute $\Pr \{ \sigma \in \Paths_{\calM}(s) \mid \sigma \models_{\calM} \Phi_1 \rU{\le k}{} \Phi_2 \}$, 
function $\checkBoundedUntil$ first populates a data structure $\mathsf{M}$ with states reachable from $s$ 
in at most $k$ steps (lines 1--28). We refer to this phase as the {\em expansion} phase\footnote{A similar approach 
is used in~\cite{HHWZ09} to analyse infinite Markov chains. However in~\cite{HHWZ09} after the expansion phase 
(that is used to compute a finite truncation of the original system), the standard model checking algorithm is used.
Indeed, differently from the solution proposed in this paper, satisfaction of $\Phi_1$ and $\Phi_2$ does not play any r\^ole.
}.

The use of this data structure enables \emph{memoization} and permits reusing the probability values computed 
already computed in different sub-formulae.
Structure $\mathsf{M}$ is a \emph{hashmap}\footnote{In this paper we use $\{ \}$ to denote the empty
hashmap, while $\textsf{M}[\mathsf{x}\mapsto \mathsf{y}]$ denotes the hashmap obtained from $\textsf{M}$ by adding the association of \emph{value} $\textsf{y}$ to \emph{key} $\mathsf{x}$. We also use $\{ \textsf{x} \mapsto \textsf{y} \}$ to denote $\{ \}[\textsf{x} \mapsto \textsf{y}]$.} that associates
each (reachable) process term $s'$ with a record of type \textsf{BURecord} with the following fields:
\begin{itemize}   
\item \textsf{term}: a value of type \textsf{proc} referring to the associated process term $s'$.
\item \textsf{prec}: a list of  \emph{predecessors} consisting of pairs $(\textsf{BURecord},\textsf{float})$. Intuitively, given  a \textsf{BURecord} $r$, a pair $(\mathsf{r}',\mathsf{p}')$ occurs in \textsf{r.prec} if and only if $\textsf{r'.term}$ evolves in one step to $\textsf{r.term}$ with probability $p'$ and has a record in $\textsf{M}$.

\item \textsf{p}: a \textsf{float} array of probabilities. The $i-th$ element in the array, \textsf{p}[i], will contain $\Pr \{ \sigma \in \Paths_{\calM}(s') \mid \sigma \models_{\calM} \Phi_1 \rU{\le i}{} \Phi_2 \}$\footnote{For the sake of readability, we explicitly consider all the components occurring in the array. When the algorithm is implemented, we do not need to store the 
whole array explicitly.}.

\item \textsf{label}: a label taking a value in $\{ \YESLabel , \NOLabel, \UNKNOWNLabel \}$. This field
takes value $\YESLabel$ when \textsf{term} satisfies $\Phi_2$. When
\textsf{term} satisfies neither $\Phi_2$ nor $\Phi_1$ the field \textsf{label} takes value $\NOLabel$ and when
it satisfies only $\Phi_1$ it takes value $\UNKNOWNLabel$.
\end{itemize}

We introduce the record precedence relation $\prec$. Let $r$ and $r'$ be two \textsf{BURecord}, we write $r\prec r'$ if and only if there exists probability $p > 0$ such that $(r,p)\in r'.\mathsf{prec}$. We will also use $r\preceq r'$ to
denote that either $r=r'$ or $r\prec r'$  and $\preceq^{i}$ for i-steps precedence.

$\checkBoundedUntil$ uses function $\createBUStructure$ to allocate new instances of \textsf{BURecord} for the starting state $s$ and further relevant states that are reachable from $s$. This function, defined in Table~\ref{tab:create_bustructure}, takes as parameter a state $s$, two
state formulas $\Phi_1$ and $\Phi_2$ and the bound $k$. The \textsf{label} field of the returned record is initialised to $\YESLabel$, 
$\NOLabel$ or $\UNKNOWNLabel$ as above by means of further calls of function $\checkDtmc$.
$\checkBoundedUntil$ initially checks the \textsf{label} for its parameter $s$ (lines 4 and 5), if such label is either $\YESLabel$ or $\NOLabel$ the values $1.0$ or $0.0$, respectively, are returned and there is no need to continue expansion. Otherwise the actual expansion phase is entered (lines 6-29). For each state to be expanded, the list of states reachable in one step from such a state is computed using function $\nxt$; a new record $r'$ is created for each state $s'$ in the list  which does not appear already in $\mathsf{M}$.  During this phase the set $\mathit{toExpand}$ is used to keep record of those $r'$ which still need to be expanded, whereas set $S_{yes}$ collects all $r'$ representing states which satisfy $\Phi_2$, i.e. are labelled $\YESLabel$. Furthermore, the list of predecessors of $r'$ is updated accordingly. Additional predecessors are added also when $r'$ was already inserted in $\mathsf{M}$ because it was visited before.
%

\begin{table}[tbp]
\begin{lstlisting}
BURecord $\createBUStructure$($s: \mathsf{proc}$ , $\Phi_1:\mathsf{formula}$ , $k: int$ , $\Phi_2: \mathsf{formula}$) {
 $l$ = $\UNKNOWNLabel$;
 p = new float[k+1];
 if ($\checkDtmc$( $s$ , $\Phi_2$ )) {
  $l$ = $\YESLabel$;
  $\forall 0\leq i\leq k. p[i]=1.0$;
 } else if ($\neg\checkDtmc$( $s$ , $\Phi_1$ )) {
  $l$ = $\NOLabel$;
 }
 return $\langle \mathsf{term}=s ; \mathsf{prec}=[] ; \mathsf{p}=p ; \mathsf{label}=l \rangle$;
}
\end{lstlisting}
\caption{\label{tab:create_bustructure} Function $\createBUStructure$}
\end{table}

When the expansion phase is completed, function $\checkBoundedUntil$ tests whether $S_{yes}$ is
empty (line 30, Table~\ref{alg:check_boundeduntil}). In that case value $0.0$ is returned because no state satisfying $\Phi_2$ 
can be reached from $s$ within $k$ steps. Hence, $\Pr \{ \sigma \in \Paths_{\calM}(s) \mid \sigma \models_{\calM} \Phi_1 \rU{\le k}{} \Phi_2 \}$ is $0.0$.  
If $S_{yes}\not=\emptyset$, function $\checkBoundedUntil$ enters the \emph{computation phase} 
(lines 32--39). This phase starts from $\YESLabel$-labelled records (now stored in variable $A$, 
indicating the \emph{active} records). 
{Then}, the probability to reach a $\YESLabel$-labelled node within 
$i$ steps is iteratively computed in a backward fashion ($i$ ranging from $1$ to $k$).  Note that a state could be the predecessor of more than one state that is on a path to a $\YESLabel$-labelled state. This is why the probability of a state to reach a $\YESLabel$-labelled state
in at most $i$ steps is the sum of the probability accumulated due to being a predecessor of other states and the probability due to being a predecessor of the currently considered state in $A$. The total probability mass is obtained when the maximal number of steps $k$ is reached. 
At the end of each iteration,  the set $A$ is updated by considering further states directly preceding those currently in $A$, i.e. $A$ is updated as follows: $\{ r | \exists r'\in A: r\preceq r' \}$.
After $i$ iterations, the set $A$ contains all the states in $\mathsf{M}$ that can reach an element in $S_{yes}$ in at most  $i$ steps.

\begin{lemma}\label{lemma:checkboundeduntil}
For each $s$, $\Phi_1$, $k$, and $\Phi_2$, let $\checkBoundedUntil(s,\Phi_1,k,\Phi_2)=p$ and $\mathsf{M}$ be
the data structure obtained at the end of the expansion phase, one of the following holds:
\begin{enumerate}
\item $\mathsf{M}[s].label=\YESLabel$ and $p=1.0$;
\item $\mathsf{M}[s].label=\NOLabel$ and $p=0.0$;
\item $\mathsf{M}[s].label=\UNKNOWNLabel$, $S_{yes}=\emptyset$ and $p=0.0$;
\item $\mathsf{M}[s].label=\UNKNOWNLabel$, $S_{yes}\not=\emptyset$ and
\[
\begin{array}{rl}
p=\Pr \{ \sigma \in \Paths_{\calM}(s) \mid &
\exists i\leq k. \mathsf{M}[\sigma[i]].label = \YESLabel \, \wedge \\
& \qquad \forall j<i. \mathsf{M}[\sigma[j]].label=\UNKNOWNLabel\}
\end{array}
\]
\end{enumerate}
\end{lemma}

\noindent
\paragraph{Proof.}
If $\checkBoundedUntil$ terminates its execution at line $4$, $5$ or $30$, then the first three
cases are readily proven. 
If $\checkBoundedUntil$ terminates
at line $40$, the last case follows directly from the fact that at line $32$ the following two loop invariants hold for iterations $i$ ranging from $1$ to $k$:\\

\noindent
$
\begin{array}{ll}
A=\{ r' | \exists r''\in S_{yes}: r'\preceq^{i} r'' \}\\[0.5em]
\forall s'.\mathsf{M}[s']=r\not=\bot , \forall j<i:\\ r.p[j]= 
\Pr \{ \sigma \in \Paths^{\calR(s,k)}_{\calM}(r.term) \mid &
\exists i'\leq j. \mathsf{M}[\sigma[i']].label = \YESLabel \, \wedge \\
&  \forall j'<i'. \mathsf{M}[\sigma[j']].label=\UNKNOWNLabel\}
\end{array}
$\\[.5em]
where $\calR(s,k)$ denotes the set of states $s'$ that are reachable from $s$ in at most $k$ steps,
while $\Paths^{\calR(s,k)}_{\calM}(s')$ denotes the set of paths starting from $s'$ that in the first
$k$ steps only pass through states in $\calR(s,k)$. Note that the following equation is straightforward:
\[
\begin{array}{rl}
\Pr \{ \sigma \in \Paths^{\calR(s,k)}_{\calM}(s) \mid &
\exists i'\leq k. \mathsf{M}[\sigma[i']].label = \YESLabel \, \wedge \\
& \qquad \forall j'<i'. \mathsf{M}[\sigma[j']].label=\UNKNOWNLabel\}=\\
\Pr \{ \sigma \in \Paths_{\calM}(s) \mid &
\exists i'\leq k. \mathsf{M}[\sigma[i']].label = \YESLabel \, \wedge \\
& \qquad \forall j'<i'. \mathsf{M}[\sigma[j']].label=\UNKNOWNLabel\}
\end{array}
\]
The proof of the Lemma follows directly from the two invariants and from the equation above.
The correctness of the invariants is proven by induction on $i$. In the following, we will use $A_{i}$ to denote the set $A$ at iteration $i$.

\noindent
\textit{Base of Induction:} If $i=1$ the statement follows directly from the fact that $A=S_{yes}\subset \calR(s,k)$.

\noindent
\textit{Induction Hypothesis:} For each $i\leq n$ we have that at line $32$ the following hold:\\[1em]
\noindent
$
\begin{array}{ll}
A_{i}  =  \{ r | \exists r''\in S_{yes}: r\preceq^{i} r''\}
\end{array}
$\\
$
\begin{array}{ll}
\forall s'.\mathsf{M}[s']=r\not=\bot ,  \forall j<i:\\ r.p[j]= 
\Pr \{ \sigma \in \Paths^{\calR(s,k)}_{\calM}(r.term) \mid &
\exists i'\leq j. \mathsf{M}[\sigma[i']].label = \YESLabel \, \wedge \\
& \forall j'<i'. \mathsf{M}[\sigma[j']].label=\UNKNOWNLabel\}
\end{array}
$

\noindent
\textit{Inductive Step:} Let us consider the case $i=n+1$. First of all, we have that: \\

$
\begin{array}{rcl}
A_{n+1} & = & \{ r  | \exists r'\in A_{n}: r\preceq r'\}\\
& \stackrel{I.H.}{=} &  \{ r | \exists r' \exists r''\in S_{yes}: r'\preceq^{n} r''\wedge r\preceq r'\} \\
& = & \{ r | \exists r''\in S_{yes}: r\preceq^{n+1} r''\}
\end{array}
$\\

\noindent
Moreover, for each $r$ such that there exists $s$: $\mathsf{M}[s]=r$ we have that  (line 35):

\[
r.p[n+1]=\sum_{\{ r'\mid r'\in A_{n} \wedge (r,p')\in r'.prec\}}r'.p[n]*p'
\]
By I.H., we have that:\\
$
\begin{array}{rl}
r'.p[n]= \Pr \{ \sigma \in \Paths^{\calR(s,k)}_{\calM}(r.term) \mid &
\exists i\leq n. \mathsf{M}[\sigma[i]].label = \YESLabel \, \wedge \\
& \forall j<i. \mathsf{M}[\sigma[j]].label=\UNKNOWNLabel\}
\end{array}
$\\
Moreover, for each $r\not\in A_{n}$, we have that $r.p[n]=0.0$, proving that  for each $r$:
\[
\begin{array}{rl}
r.p[n+1]= \Pr \{ \sigma \in \Paths^{\calR(s,k)}_{\calM}(r.term) \mid &
\exists i\leq n+1. \mathsf{M}[\sigma[i]].label = \YESLabel \, \wedge \\
& \forall j<i. \mathsf{M}[\sigma[j]].label=\UNKNOWNLabel\}
\end{array}
\]
which proves the correctness of the two invariants.

{\hfill $\Box$}

%


\subsection{Computing Unbounded Until Probability}

In this section we present function $\checkUnBoundedUntil$ that can be used to compute
the probability of the set of paths satisfying $\Phi_1 \rU{}{} \Phi_2$ starting from a state $s$.
Similarly to function $\checkBoundedUntil$ considered in the previous section, function 
$\checkUnBoundedUntil$, defined in Table~\ref{alg:check_unboundeduntil}, is structured in two phases: an 
\emph{expansion phase} (lines 2--30) and a \emph{computation phase} (lines 35-48). 
In the first phase, all the process terms reachable from state $s$, that are relevant for
the computation of $\Pr \{ \sigma \in \Paths_{\calM}(s) \mid \sigma \models_{\calM} \Phi_1 \rU{}{} \Phi_2 \}$,
are generated. The discovered terms are stored in a hash map $\mathsf{M}$ associating each reachable process term 
with an instance of record \textsf{UURecord}. This record type is the same as \textsf{BURecord} except for the field \textsf{p} which
is replaced by two arrays of float elements $p_{yes}$ and $p_{no}$. The role of these two fields will be clarified soon.
Function $\createUStructure$, defined in Table~\ref{tab:create_ustructure}, is used to allocate a new instance of 
\textsf{UURecord}. 
%

We use the same notation for the record precedence relation $\prec$ on \textsf{UURecord} as was introduced for \textsf{BURecord}. 
We will also use $\preceq^{*}$ to denote the transitive closure of $\preceq$. 
During the \emph{expansion phase} (see Table~\ref{alg:check_unboundeduntil}, lines 9-30) 
the sets $S_{yes}$ and $S_{no}$ are populated. These sets eventually contain all the records
that are labelled $\YESLabel$ and $\NOLabel$, respectively. The expansion terminates when no new 
record is found. 
\begin{table}[tbp]
\begin{lstlisting}
UURecord $\createUStructure$($s: \mathsf{proc}$ , $\Phi_1:\mathsf{formula}$ , $\Phi_2: \mathsf{formula}$) {
 $l$ = $\UNKNOWNLabel$;
 $p_{yes}$ = new float[2];
 $p_{no}$ = new float[2];
 if ($\checkDtmc$( $s$ , $\Phi_2$ )) {
  $l$ = $\YESLabel$;
  $p_{yes}$ = { 1.0 , 1.0 };
  $p_{no}$ = { 0.0 , 0.0 };
 } else if ($\neg\checkDtmc$( $s$ , $\Phi_1$ )) {
  $l$ = $\NOLabel$;
  $p_{yes}$ = { 0.0 , 0.0 };
  $p_{no}$ = { 1.0 , 1.0 };
 }
 return $\langle \mathsf{term}=s ; \mathsf{prec}=[] ; \mathsf{p}_{yes}=p_{yes} ; \mathsf{p}_{no}=p_{no} ; \mathsf{label}=l \rangle$;
}
\end{lstlisting}
\caption{\label{tab:create_ustructure} Function $\createUStructure$}
\end{table}
Before starting the \emph{computation phase}, it is first checked whether $S_{yes}$ is empty (Line 31 in Table~\ref{alg:check_unboundeduntil}) .
If $S_{yes}$ is empty then the value $0.0$ is returned because no state satisfying $\Phi_2$ can be reached starting from $s$.
If $S_{yes}$ is not empty, all the records that \emph{cannot reach}  $\YESLabel$-labelled records are added to $S_{no}$, their labels updated to $\NOLabel$ and the probability $p_{no}$ set to 1.
%
If the resulting set $S_{no}$ at this point is empty, the value $1.0$ is returned because in this case $s$ can only eventually reach $\YESLabel$-labelled records.
If $S_{no}\not=\emptyset$, all the records that \emph{cannot reach} $\NOLabel$-labelled records 
are added to $S_{yes}$, labelled by $\YESLabel$ and their probability value is set to 1.0. An example could be the occurrence of bottom strongly connected components (BSCC) consisting exclusively of states satisfying $\Phi_1$. The states on such BSCCs cannot reach states that are labelled $\YESLabel$ (or $\NOLabel$), but they can be treated as states labelled $\NOLabel$.  

The \emph{computation phase} (starting at line $39$ in Table~\ref{alg:check_unboundeduntil})  operates on a set of \emph{active} records $A$. Initially, $A$
is  $S_{yes}\cup S_{no}$. At the end of each iteration (line 42 in Table~\ref{alg:check_unboundeduntil})  $A$ is extended with all the process terms that are able to reach  elements in $A$ in one step.
At the end of iteration $i$, for each element $r$ in $A$, $\mathsf{r.p_{yes}}[ i \mod 2]$ contains the probability mass of the set of paths
starting from $\mathsf{r.term}$ that reach within $i$ steps a $\YESLabel$-labelled term while passing only through $\UNKNOWNLabel$-labelled records.
Similarly, at the end of the same iteration $i$,  $\mathsf{r.p_{no}}[i \mod 2]$ stores the probability of the set of paths starting from $\mathsf{r.term}$ that reach within 
$i$ steps a $\NOLabel$-labelled term while only passing through $\UNKNOWNLabel$-labelled records.
The computation phase terminates when, for each record $r$ stored in $\mathsf{M}$, the following holds: 
$\mathsf{r.p_{yes}}[ i \mod 2]+\mathsf{r.p_{no}}[i \mod 2] \geq 1-\varepsilon$, where $\varepsilon$ is a given accuracy level. 
The reason for computing both $\mathsf{r.p_{yes}}$ and $\mathsf{r.p_{no}}$ is that this way a well-known property of  transient DTMCs can be exploited to detect and predict when a sufficiently accurate result has been obtained and the computation can be terminated (see Sect.~\ref{sec:complex}). Furthermore, only the current and next value of the probability needs to be stored. This is obtained by using two fields that are addressed via the index modulo 2. For the rest, the computation of the respective probabilities of $p_{yes}$ and $p_{no}$ follow the same pattern as for bounded until.
%
%

\begin{table}[t!]
\begin{lstlisting}
float $\checkUnBoundedUntil$($s: \mathsf{proc}$ , $\Phi_1:\mathsf{formula}$ , $\Phi_2: \mathsf{formula}$) {
 r = $\createUStructure$( s , $\Phi_1$ , $\Phi_2$ );
 $M$ = $\{s \mapsto r\}$;
 if (r.label == $\YESLabel$) { return 1.0; }
 if (r.label == $\NOLabel$) { return 0.0; }
 $S_{yes}$ = $\emptyset$;
 $S_{no}$ = $\emptyset$;
 $toExpand = \{ r \}$;
 while $(toExpand \not= \emptyset)$ {
  $T=toExpand$;
  $toExpand = \emptyset$;
  for ($r\in T$) {
   $lst$ = $\nxt(r.term)$;
   for $(s',p')\in lst$ {
    $r'$=$M[s']$;
    if ($r'==\bot$) {
      $r'$ = $\createUStructure$( $s'$ , $\Phi_1$ , $\Phi_2$ );
      $M$ = $M[s' \mapsto r']$;
      if ($r'.label$ == $\YESLabel$) { 
       $S_{yes}$= $S_{yes}\cup \{ r' \};$ 
      } else if ($r'.label$ == $\NOLabel$) {
       $S_{no}$=$S_{no}\cup \{ r' \};$
      } else {
       $toExpand$ = $toExpand\cup \{ r' \}$; 
      }
    }
    $r'.prec$ = $(r,p)::r'.prec$;
   }
  }
 }
 if ($S_{yes}==\emptyset$) { return 0.0; }
 $S_{no}= \{ r | \not\exists r'\in S_{yes}. r \preceq^{*} r' \}$
 $\forall r\in S_{no}. r.p_{no} =\{ 1 , 1 \}$, $r.label=\NOLabel$;
 if ($S_{no}==\emptyset$) { return 1.0; }
 $S_{yes}= \{ r | \not\exists r'\in S_{no}. r \preceq^{*} r' \}$ 
 $\forall r\in S_{yes}. r.p_{yes} =\{ 1 , 1 \}$, $r.label=\YESLabel$;
 $A$ = $S_{yes}\cup S_{no}$;
 $i$ = $0$;
 while ($\exists (s,r)\in M: r.p_{yes}[i \mod 2]+r.p_{no}[i \mod 2]<1-\varepsilon$) {
   $\forall r\in A. r.p_{yes}[(i+1) \mod 2 ])=0$;
   $\forall r\in A. r.p_{no}[(i+1) \mod 2 ])=0$;
  for ($r\in A$) {
   for ($(r',p')\in r.prec$) {
    $r'.p_{yes}[(i+1) \mod 2 ]$ = $r'.p_{yes}[(i+1) \mod 2]+p'*r.p_{yes}[i \mod 2]$;   
    $r'.p_{no}[(i+1) \mod 2 ]$ = $r'.p_{no}[(i+1) \mod 2]+p'*r.p_{no}[i \mod 2]$;   
   }
  }
  $i$ = $i+1$;
  $A$ = $\{ r | \exists r'\in A: r\preceq r' \}$;
 }
 return $r.p[i\mod 2]$;
}
\end{lstlisting}
\caption{\label{alg:check_unboundeduntil} Function $\checkUnBoundedUntil$}
\end{table}

\begin{lemma}\label{lemma:checkunboundeduntil}
For each $s$, $\Phi_1$ and $\Phi_2$, let $\checkUnBoundedUntil(s,\Phi_1,\Phi_2)=p$ and $\mathsf{M}$ be
the data structure obtained at the end of the expansion phase, one of the following holds:
%
\begin{enumerate}
\item $\mathsf{M}[s].label=\YESLabel$ and $p=1.0$;
\item $\mathsf{M}[s].label=\NOLabel$ and $p=0.0$;
\item $\mathsf{M}[s].label=\UNKNOWNLabel$, $S_{yes}=\emptyset$ and $p=0.0$;
\item $\mathsf{M}[s].label=\UNKNOWNLabel$, $S_{no}=\emptyset$ and $p=1.0$;
\item $\mathsf{M}[s].label=\UNKNOWNLabel$, $S_{yes},S_{no}\not=\emptyset$ and
\begin{center}
$
\begin{array}{rl}
|\Pr \{ \sigma \in \Paths_{\calM}(s) \mid &
\exists i. \mathsf{M}[\sigma[i]].label = \YESLabel \, \wedge \\
& \qquad \forall j<i. \mathsf{M}[\sigma[j]].label=\UNKNOWNLabel\}-p|\leq \varepsilon
\end{array}
$
\end{center}
\end{enumerate}
\end{lemma}

\paragraph{Proof.}
If $\checkUnBoundedUntil$ terminates its computation at line $4$, $5$, $31$ or $34$, then the first four
cases apply respectively in a straightforward way. 
If $\checkUnBoundedUntil$ terminates
at line $51$, the statement follows directly from the fact that the following three invariants hold at line $39$ for iteration $i$:\\[.15cm]
\noindent
$
A=\{ r' | \exists r''\in S_{yes}\cup S_{no}: r' \preceq^{i} r'' \}
$\\
%
%
\noindent
$
\begin{array}{ll}
\forall s'.\mathsf{M}[s']=r\not=\bot ,\\
r.p_{yes}[i \mod 2] = 
\Pr \{ \sigma \in \Paths_{\calM}(r.term) \mid &
\exists i'\leq i. \mathsf{M}[\sigma[i']].label = \YESLabel \, \wedge \\
& \forall j'<i'. \mathsf{M}[\sigma[j']].label=\UNKNOWNLabel\}\\[.15cm]
%
%
\forall s'.\mathsf{M}[s']=r\not=\bot ,\\ 
r.p_{no}[i \mod 2] = 
\Pr \{ \sigma \in \Paths_{\calM}(r.term) \mid &
\exists i'\leq i. \mathsf{M}[\sigma[i']].label = \NOLabel \, \wedge \\
& \forall j'<i'. \mathsf{M}[\sigma[j']].label=\UNKNOWNLabel\}
\end{array}
$

Let $\checkUnBoundedUntil(s,\Phi_1,\Phi_2)=p$ and $r=\mathsf{M}[s]$, then $p=r.p_{yes}[i\mod 2]$ and
%
$1 \geq r.p_{yes}[i\mod 2]+r.p_{no}[i\mod 2]\geq 1-\varepsilon=(p^{s}_{yes}+p^{s}_{no})-\varepsilon$
%
where:\\[.15cm]
$
\begin{array}{rl}
p^{s}_{yes}  = 
\Pr \{ \sigma \in \Paths_{\calM}(r.term) \mid &
\exists i: \mathsf{M}[\sigma[i]].label = \YESLabel \, \wedge \\
&  \forall j<i. \mathsf{M}[\sigma[j]].label=\UNKNOWNLabel\}\\[.15cm]
p^{s}_{no}  = 
\Pr \{ \sigma \in \Paths_{\calM}(r.term) \mid &
\exists i. \mathsf{M}[\sigma[i]].label = \NOLabel \, \wedge \\
&  \forall j<i. \mathsf{M}[\sigma[j]].label=\UNKNOWNLabel\}
\end{array}
$\\
Since $r.p_{yes}[i\mod 2]\leq p^{s}_{yes}$ and $r.p_{no}[i\mod 2]\leq p^{s}_{no}$, this implies that:
%
\[
(p^{s}_{yes}-r.p_{yes}[i\mod 2]) + (p^{s}_{no} - r.p_{no}[i\mod 2]) - \varepsilon \leq 0
\]
Hence:
$
p^{s}_{yes}-r.p_{yes}[i\mod 2] \leq  \varepsilon
$
which proves Lemma~\ref{lemma:checkunboundeduntil}, apart from the three invariants, considered above, which are proven below.  The proof of the first invariant is identical to the one considered in 
the proof of Lemma~\ref{lemma:checkboundeduntil} and we omit it. The other two invariants are proven by induction on 
$i$. 
\ifTR
We only show the proof for the invariant concerning $p_{yes}$, the other being very similar.
\fi

\noindent
\textit{Base of Induction:} If $i=1$ the statement follows directly from the fact that $A_{0}=S_{yes}\cup S_{no}$, where we
use $A_{i}$ to denote the value of set $A$ at iteration $i$. Note that for each $r\in M$, if $r\not\in A_{0}$, $r.p_{yes}[i\mod 2]=r.p_{no}[i\mod 2]=0.0$.

\noindent
\textit{Induction Hypothesis:} For each $i\leq n$ we have that at line $39$ the following holds:
\[
\begin{array}{ll}
\forall s'.\mathsf{M}[s']=r\not=\bot , \\ 
r.p_{yes}[i \mod 2] = 
\Pr \{ \sigma \in \Paths_{\calM}(r.term) \mid &
\exists i'\leq i. \mathsf{M}[\sigma[i']].label = \YESLabel \, \wedge \\
& \forall j'<i'. \mathsf{M}[\sigma[j']].label=\UNKNOWNLabel\}
\end{array}
\]
\ifTR
\[
\begin{array}{ll}
\forall s'.\mathsf{M}[s']=r\not=\bot , \\
r.p_{no}[i \mod 2] = 
\Pr \{ \sigma \in \Paths_{\calM}(r.term) \mid &
\exists i'\leq i. \mathsf{M}[\sigma[i']].label = \NOLabel \, \wedge \\
&  \forall j'<i'. \mathsf{M}[\sigma[j']].label=\UNKNOWNLabel\}
\end{array}
\]
\fi

\noindent
\textit{Inductive Step:} Let us consider the case $i=n+1$.  For each $r$ such that there exists $s$: $\mathsf{M}[s]=r$ we have that (lines 44-45 in Table~\ref{alg:check_unboundeduntil}):
\[
r.p_{yes}[(n+1)\mod 2]=\sum_{\{ r'\mid r'\in A_{n} \wedge (r,p')\in r'.prec\}}r'.p_{yes}[n\mod 2]*p'
\]
\ifTR
\[
r.p_{no}[(n+1)\mod 2]=\sum_{\{ r'\mid r'\in A_{n} \wedge (r,p')\in r'.prec\}}r'.p_{no}[n\mod 2]*p'
\]
\fi
By induction hypothesis, we have that:
\[
\begin{array}{rl}
r'.p_{yes}[n\mod 2]= \Pr \{ \sigma \in \Paths_{\calM}(r.term) \mid &
\exists i\leq n. \mathsf{M}[\sigma[i]].label = \YESLabel \, \wedge \\
&  \forall j<i. \mathsf{M}[\sigma[j]].label=\UNKNOWNLabel\}
\end{array}
\]
\ifTR
\[
\begin{array}{rl}
r'.p_{no}[n\mod 2]= \Pr \{ \sigma \in \Paths_{\calM}(r.term) \mid &
\exists i\leq n. \mathsf{M}[\sigma[i]].label = \NOLabel \, \wedge \\
& \forall j<i. \mathsf{M}[\sigma[j]].label=\UNKNOWNLabel\}
\end{array}
\]
\fi
Since for each $r'\in \mathsf{M}$, such that $r'\not\in A_{n}$, $r'.p_{yes}=r'.p_{no}=\{ 0 , 0 \}$, then 
\[
\begin{array}{rl}
r.p_{yes}[n+1]= \Pr \{ \sigma \in \Paths_{\calM}(r.term) \mid &
\exists i\leq n+1. \mathsf{M}[\sigma[i]].label = \YESLabel \, \wedge \\
& \forall j<i. \mathsf{M}[\sigma[j]].label=\UNKNOWNLabel\}
\end{array}
\]
\ifTR
\[
\begin{array}{rl}
r.p_{no}[n+1]= \Pr \{ \sigma \in \Paths_{\calM}(r.term) \mid &
\exists i\leq n+1. \mathsf{M}[\sigma[i]].label = \NOLabel \, \wedge \\
& \forall j<i. \mathsf{M}[\sigma[j]].label=\UNKNOWNLabel\}
\end{array}
\]
\fi
\ifTR
which proves the two invariants.
\fi
{\hfill $\Box$}

\subsection{Termination and Complexity}
\label{sec:complex}
The idea of computing both the probability of the set of paths that satisfy a path-formula and the probability of the set of those that do not is motivated by the possibility to exploit an interesting property of {\em transient} DTMCs~\cite{Kemeny1976}. A Markov chain is transient iff all its recurrent states are absorbing. This is the case for the DTMCs that are constructed by the model checking algorithm, in particular the states labelled $\YESLabel$ and $\NOLabel$ are absorbing and the states labelled $\UNKNOWNLabel$ are transient.
The probability matrix $\mathbf{P}$ of a generic transient DTMC can be arranged as:
 \begin{equation}
\label{eq:transition_matrix_transient_chain}
\begin{footnotesize}
\begin{array}{ccc}
\mathbf{P} & = &
\begin{array}{cc}
 & E \qquad \tilde{E} \\
 \begin{array}{c}
	E \\ \tilde{E}
 \end{array} & \left(\begin{array}{c}
 I \qquad 0 \\
 R \qquad Q
 \end{array}
 \right)
\end{array}
\end{array}
\end{footnotesize}
\end{equation}
where $E$ and $\tilde{E}$ denote the set of recurrent states and the transient states of the DTMC, respectively. For this kind of Markov chains the following lemma~\cite[pag. 107]{Kemeny1976} can be easily shown to hold using an inductive argument:

\begin{lemma}
\label{lemma:transient_chain}
Let $\mathcal{D} = \defdtmc $ be a transient DTMC, with $\mathbf{P}$ of the form shown in~(\ref{eq:transition_matrix_transient_chain}), then
$\displaystyle \lim_{i \to \infty} Q^i = 0$ and:
\begin{center}$
\begin{footnotesize}
\begin{array}{ccc}
\mathbf{P}^i & = &
\begin{array}{ccc}
\begin{array}{c}
  \\
  \begin{array}{c}
 	E \\ \tilde{E}
  \end{array} 
\end{array}
&
\begin{array}{c}
E \\
  \left(
  \begin{array}{c}
 	I \\ (I + Q + \cdots + Q^{i-1})R
  \end{array} 
  \right.
\end{array}
&
\begin{array}{c}
\tilde{E} \\
  \left.
  \begin{array}{c}
 	0 \\ Q^i
  \end{array} 
  \right)
\end{array}
\end{array}
\end{array}
\end{footnotesize}
$\end{center}
%
\qed
\end{lemma}

\begin{lemma}\label{lemma:checkunboundeduntiltermination}
Let $s$ be such that the set $\{ s' | \exists k. s'\in \calR(s,k)\}$ of states reachable from $s$ is finite; then, for each $\Phi_1$ and $\Phi_2$, 
$\checkUnBoundedUntil(s,\Phi_1,\Phi_2)$ terminates.
\end{lemma}

\paragraph{Proof}
The statement follows directly from Lemma~\ref{lemma:transient_chain} by observing that two transient DTMCs are implicitly considered in  function $\checkUnBoundedUntil$. One in which $E=S_{yes}$
while $\tilde{E}$ consists of the set of records labelled $\UNKNOWNLabel$; this DTMC is used to compute 
the probability mass of the set of paths {\em satisfying} $\Phi_1\rU{} \Phi_2$. The other transient DTMC,
is the one where $E$ is $S_{no}$ while $\tilde{E}$ is again the set of records labelled $\UNKNOWNLabel$. This
second DTMC is used to compute the probability mass of the set of paths that do {\em not} satisfy $\Phi_1\rU{} \Phi_2$.
Function $\checkUnBoundedUntil$ terminates when the sum of the two computed probability values 
differs from $1.0$ by less than a given accuracy bound $\varepsilon$. Note that this difference is in fact the 
\emph{total remaining probability} in $Q^{i}$ (where $i$ is the current iteration). Lemma~\ref{lemma:transient_chain}
guarantees that the threshold is eventually reached.
\qed

Lemma~\ref{lemma:checkunboundeduntiltermination} guarantees that the algorithm always terminates when the 
set of states reachable from $s$ is finite. For what concerns complexity, the number of iterations that is required to complete the computation
depends on the accuracy bound $\varepsilon$ and on the \emph{stiffness} of the model. In particular, the number of iterations
is bounded by $\frac{\log{\varepsilon}}{\log(\max_{i}\{\sum_{j}Q_{i,j}\})}$.
%
%

\section{On-the-fly Model checking: a Preliminary Assessment}
\label{sec:atwork}

A prototype of the model checking algorithm has been implemented in Java, together with an interpreter of the PRISM language~\cite{KNP04}, and used for a first comparison of its performance with that of PRISM\footnote{Details of the 
experiments are available at \url{http://j-sam.sourceforge.net}.}. The choice of this specific probabilistic model-checker is justified by the fact that it is the state of the art for what concerns advanced state space reduction techniques.


The first case study we consider is the self-stabilising algorithm of Herman~\cite{Her90,KNP04}. This algorithm 
defines a protocol for a network of processes arranged in a ring. Each process can be either \emph{active}
or \emph{passive}. A configuration is \emph{stable} when only one process is active. The
algorithm guarantees that, when starting from an unstable configuration, the  system
is able to return to a stable configuration with probability 1 within a finite number of steps. We assume an initial
configuration where all the processes are active and four properties that require the full state space to be expanded: (\textbf{P1}) the probability to reach a stable configuration within $50$ steps is greater than $p$; (\textbf{P2}) the probability to reach eventually a stable configuration is greater than $p$; 
(\textbf{P3}) the probability to reach a stable configuration within $50$ steps while the first process remains active
is greater than $p$; (\textbf{P4}) the probability to reach eventually a stable configuration while the first process remains
active is greater than $p$. The model checking time (in milliseconds if not specified otherwise) needed to perform these analyses is reported in Table~\ref{tab:results} (left), where $N$ indicates the number of processes in the ring, O the on-the-fly model-checker and P the PRISM model-checker.
\begin{table}[tbp]
\begin{center}
{\footnotesize
\begin{tabular}{|c|r|r|r|r|r|r|r|r|} \hline
Property:& \multicolumn{2}{|c|}{\textbf{P1}} & \multicolumn{2}{|c|}{\textbf{P2}} & \multicolumn{2}{|c|}{\textbf{P3}} & \multicolumn{2}{|c|}{\textbf{P4}} \\ \hline
$N$	& \quad O & P	        & \quad O & P 	    & \quad O   & P	     & \quad O       & P \\ \hline
$3$	& $1$	& $2$	& $1$	& $2$   & $1$	       & $2$	     & $1$             & $2$ \\
$5$	& $2$	& $3$	& $2$	& $3$   & $2$          & $3$	      & $2$            & $3$ \\
$7$	& $3$	& $4$	& $2$	& $3$   & $2$	       & $4$    &  $3$            & $4$ \\
$9$	& $8$	& $8$     & $4$      & $4$ & $4$        & $6$     & $4$          & $6$ \\ \hline
\end{tabular}\qquad
\begin{tabular}{|c|r|r|r|r|} \hline
 Property:& \multicolumn{2}{|c|}{\textbf{P1}} & \multicolumn{2}{|c|}{\textbf{P2}} \\ \hline
$N$	& \quad O      & P      & \quad O & P	\\ \hline
$3$  & $1$   & $3$       & $1$ & $3$ \\
$5$	& $1$  & $7$       & $1$ & $12$ \\
$9$	& $1$  & $402$     & $2$ & $711$ \\
$11$ & $1$  & $6.91s$  & $2$ & $8.027s$ \\	
$15$ & $1$  & -             & $2$ & - \\
$21$ & $2$  & -             & $2$ & - \\ \hline
\end{tabular}
}
\end{center}
\caption{\label{tab:results} MC time of Self-stabilisation (left) and Dining Philisophers (right) }
\end{table}
These first results show that the efficiency of the on-the-fly algorithm is \emph{comparable} with that of PRISM for PCTL\footnote{Experiments have been performed with an Intel Core i7 $1.7GHz$, RAM $8Gb$.  For PRISM the model generation time has not been considered.}.   
The efficiency of the  on-the-fly algorithm increases drastically when, to verify a given property, only
a subset of the state space is needed. This is the case for the second case study where a probabilistic
variant of the \emph{Dining Philosophers} is considered. We consider two properties: (\textbf{P1}) 
Philosopher $1$ is the first to eat within the next $20$ steps with probability greater than $p$; 
(\textbf{P2}) With probability greater than $p$, philosopher $1$ is the first to eat\footnote{Properties of single objects are very relevant in e.g. population models~\cite{La+13a}.}. The execution times of
both on-the-fly model checking and PRISM are reported in Table~\ref{tab:results} (right).
%
%
Note that for these specific properties the on-the-fly model-checker is able
to give a result within a few milliseconds even for a system composed of $15$ or $21$ philosophers. PRISM instead raises an \emph{out-of-memory} exception for these cases. Which of the two model checkers is more convenient to use depends on the problem at hand. In highly symmetric cases PRISM is expected to perform better because it can exploit powerful reduction techniques based on the underlying MTBDD structure. Having all techniques available gives the greatest advantage. 

\section{Conclusions and Future Work}
\label{CONC:FW}

In this paper we have presented an innovative {\em local, on-the-fly} PCTL model checking approach including both bounded and unbounded modalities. The model checking algorithm is parametric w.r.t. the language and the specific semantic model of interest. The algorithm for unbounded until is new and exploiting a well-known property of transient DTMCs to obtain an efficient procedure to compute the probability of unbounded path formulas with a desired accuracy. Correctness proofs of the algorithms have been provided and a prototype implementation with the PRISM language as front-end has been used to perform a comparison of its efficiency both in memory use and in time and for exact probabilistic model checking. The model checker has also been instantiated and used for fast on-the-fly PCTL model checking
for discrete time synchronous population models obtaining a scalability independent of the size of the population~\cite{La+13a,LLM13b}
%
%
Further work is planned on extensions that concern spatial aspects of systems as well as the application to a larger range of case-studies incorporating further probabilistic languages and related semantics. 
%
A more detailed comparison with statistical model checking, e.g. \cite{Younes2004}, or partial order reduction techniques, e.g. \cite{BDG06}, is also planned.

\bibliographystyle{eptcs}
\bibliography{abbr,opmc,xref}

\end{document}